\documentclass{article}
\usepackage{url}
\usepackage[sort,numbers]{natbib}
\usepackage{color}
\usepackage{spconf,amsmath,graphicx}
\usepackage{multirow}
\usepackage{makecell}
\usepackage{booktabs}
\usepackage{amssymb}
\usepackage{amsmath}
\usepackage{graphicx}
\usepackage{enumitem}
\usepackage{threeparttable}
\usepackage{diagbox}
\usepackage{setspace}
\usepackage{array,caption}
\usepackage[labelfont=bf]{caption}

\title{\MakeUppercase{SSVMR: Saliency-Based self-training for video-music retrieval}}
\name{Xuxin Cheng, Zhihong Zhu, Hongxiang Li, Yaowei Li, Yuexian Zou$^{*}$\thanks{
*Corresponding author
}
}
\address{
   ADSPLAB, School of ECE, Peking University\\
\{chengxx, zhihongzhu, lihongxiang, ywl\}@stu.pku.edu.cn,  zouyx@pku.edu.cn
}

\usepackage{hyperref}
\begin{document}
\topmargin=0mm
%
\maketitle
\begin{abstract}
With the rise of short videos, the demand for selecting appropriate \textit{background music}~(BGM) for a video has increased significantly, \textit{video-music retrieval}~(VMR) task gradually draws much attention by research community. As other cross-modal learning tasks, existing VMR approaches usually attempt to measure the similarity between the video and music in the feature space. However, they (1) neglect the inevitable label noise; (2) neglect to enhance the ability to capture critical video clips. In this paper, we propose a novel saliency-based self-training framework, which is termed SSVMR. Specifically, we first explore to fully make use of the information containing in the training dataset by applying a semi-supervised method to suppress the adverse impact of label noise problem, where a self-training approach is adopted. In addition, we propose to capture the saliency of the video by mixing two videos at span level and preserving the locality of the two original videos. Inspired by back translation in NLP, we also conduct back retrieval to obtain more training data. Experimental results on MVD dataset show that our SSVMR achieves the state-of-the-art performance by a large margin, obtaining a relative improvement of 34.8\% over the previous best model in terms of R@1.
\end{abstract}

%
\begin{keywords}
Video-music Retrieval, Cross-modal Matching, Self-training, Label Noise
\end{keywords}
\section{Introduction}
\vspace{-0.5em}
\label{sec:intro}
\textit{Background music}~(BGM) can adjust atmosphere, enhance emotional expression of the video, and achieve an immersive feeling for the audience. With the rise of short videos, the demand of selecting a proper BGM for a video has increased significantly. However, accomplishing this work manually requires a great deal of expertise. To handle this problem, \textit{video-music retrieval}~(VMR) has received increasing attention from researchers. VMR aims to match the input video with the appropriate BGM from the audio library.

Recently, existing VMR approaches attempt to measure the similarity between the video and music in the feature space. As other cross-model tasks~\cite{ye2022cross,li2023generating,tian2022bayes}, some conventional VMR methods conduct cross-modal matching leveraging external data, such as metadata~\cite{brochu2003sound, KiHoShin2017MusicSW,BochenLi2019QueryBV}. Metadata typically provides key information about the content of videos or music in a \textit{quick and simple} way, which makes multimedia content as searchable as text. \cite{brochu2003sound} uses color histogram representations of album covers to index music pieces. \cite{KiHoShin2017MusicSW} retrieves BGMs for videos by measuring similarity of their emotion tags. \cite{BochenLi2019QueryBV} leverages their emotion tags as joint constraints to pre-train music and video feature extractors, which can better align the embeddings of video and music modalities. Unfortunately, metadata is not always available, especially when the size of data is large. More recently, some work has attempted to accomplish VMR tasks using a content-based approach without the use of metadata. \cite{hong2018cbvmr} introduces a two-branch deep network that associates videos and music considering inter and intra-modal relationships. \cite{JingYi2021CrossmodalVA} proposes a cross-generation strategy which can also better align latent embedding of videos and music. 

Despite the promising progress that existing VMR models have achieved, we discover they suffer from two main issues:

\noindent \textbf{(1) Neglecting the inevitable label noise in VMR~\cite{frenay2013classification}}. Label noise can easily result in overfitting of the model~\cite{ShengLiu2020EarlyLearningRP, zhu2022dynamic}. Compared to other cross-modal counterparts, VMR task is likely to have more label noise for two reasons: a) The labels of VMR dataset are subjective, and different experts may choose very different BGMs for the same video; b) VMR dataset HIMV-50K~\cite{9533662} and MVD~\cite{AlexanderSchindler2016HarnessingMV} both consist of three kinds of \textit{video-music} pair, including official music videos, parody music videos and user-generated videos, resulting in some variation in the relationship between data and labels.

\noindent \textbf{(2) Neglecting to enhance the ability to capture critical video clips.} Intuitively, different clips from the video play different roles in determining the corresponding music. There may be a small segment of the video that plays the most important role. However, previous work treats every video clip equally, which may lead to a decrease in model performance.


\begin{figure*}[t]
  \centering
  \includegraphics[width=\linewidth]{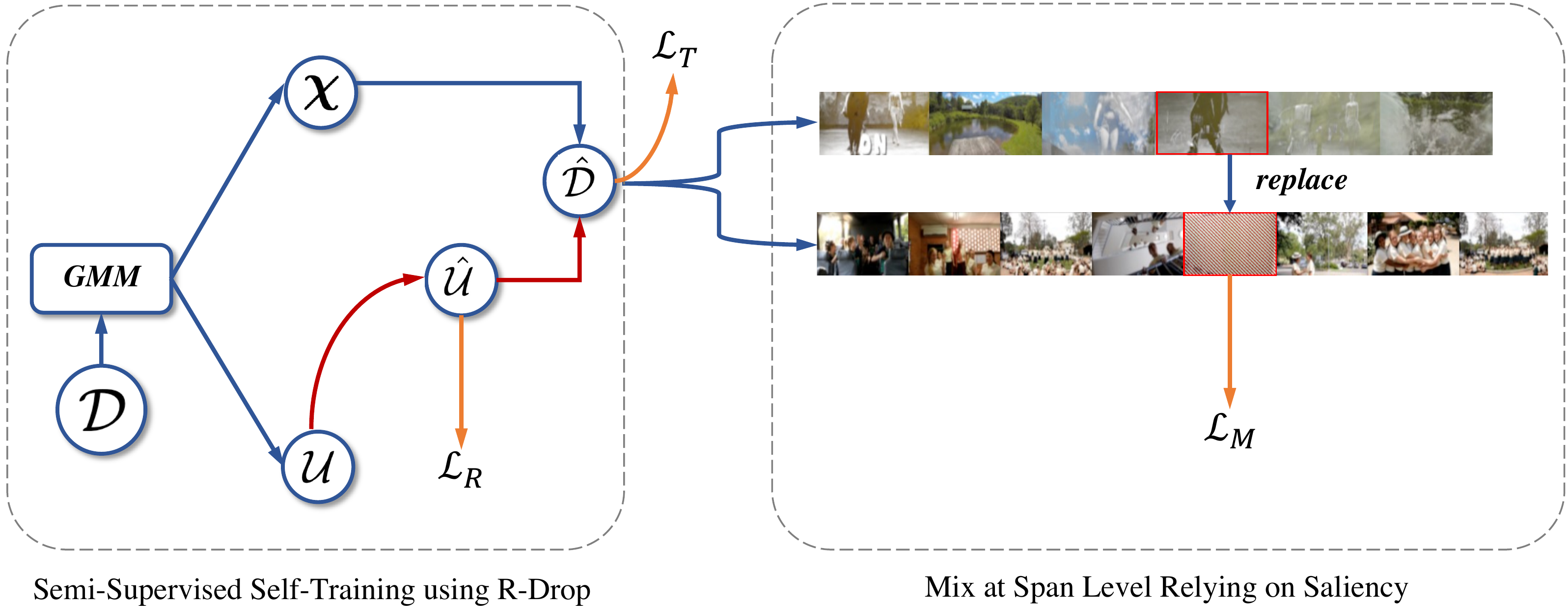}
  \caption{The architecture of our proposed framework, which consists of two components: (1) Semi-Supervised Self-Training using R-Drop; (2) Mix at Span Level Relying on Saliency. $\mathcal{D}$ denotes original \textit{video-music} pairs, including a labeled set $\mathcal{X}$ and an unlabeled set $\mathcal{U}$. $\hat{\mathcal{U}}$ denotes $\mathcal{U}$ with new soft labels. $\hat{\mathcal{D}}$ denotes the combination of $\mathcal{X}$ and $\hat{\mathcal{U}}$.}
  \label{fig:network}
\vspace{-1em}
\end{figure*}


In this paper, we propose \textbf{S}aliency-based \textbf{S}elf-training for \textbf{V}ideo-\textbf{M}usic \textbf{R}etrieval (\textbf{SSVMR}) to tackle the above two issues. For the first issue, we use Gaussian Mixture Model (GMM) to select the samples which are more likely to be wrong and replace their original labels by soft labels through sharpen. Then, we introduce R-Drop~\cite{wu2021r} loss which regularizes two predicted distributions obtained by inputting the same sample twice to reduce the confirmation bias in self-training. For the second issue, we mix the two videos at span level relying on saliency. By doing so, more important information related to the prediction will be retained and the locality of the two original videos will be preserved. Moreover, inspired by back translation~\cite{sennrich2016improving} in NLP, we conduct back retrieval from music-video retrieval model to enlarge the dataset. Following~\cite{9533662}, we train our model on HIMV-50K~\cite{9533662} dataset and evaluate it on MVD~\cite{AlexanderSchindler2016HarnessingMV} dataset. Experiments and further analysis show that SSVMR can effectively handle label noise and improve the ability of model to capture critical video clips, which achieves performance over previous state-of-the-art methods by a large margin. To best of our knowledge, our framework is the first to handle label noise and consider saliency for VMR task.


\section{Method}
\vspace{-0.5em}
In this section, we first introduce the problem formulation of VMR in Section \ref{sec:2.1} and the backbone in Section \ref{sec:2.2}, then describe the details of our framework. As shown in Fig.\ref{fig:network}, it mainly consists of self-training with R-Drop module in Section \ref{sec:2.3}, mix at span level relying on saliency module in Section \ref{sec:2.4} and back retrieval module in Section \ref{sec:2.5}.
\label{sec:pro}
\vspace{-1em}
\subsection{Problem Formulation}
\label{sec:2.1}
\vspace{-0.5em}

The dataset for VMR task usually contains \textit{video-music} pairs, which can be denoted as $\mathcal{D}=\{(v, m)\}$, where $v$ is the embedded features of the video, and $m$ is the corresponding embedded features of the BGM. For a set of video  $\mathcal{V}$ and a set of music  $\mathcal{M}$, a VMR method aims to retrieve a list of music candidates  $\mathcal{C}(m) \subset \mathcal{M}$  where each music  $m$ in $\mathcal{C}(m)$ is a potential match for the query video $v$ from $\mathcal{V}$.
\vspace{-1em}
\subsection{Backbone}
\label{sec:2.2}
\vspace{-0.5em}
We use CBVMR~\cite{hong2018cbvmr} as the backbone, which uses a two-branch neural network and combines the inter-modal ranking constraint with soft intra-modal structure constraint to construct an unified embedding network. It tries to minimize the empirical risk by applying the triplet loss $\mathcal{L}_{T}$ as follows:
\begin{equation}
\setlength{\abovedisplayskip}{3pt}
\setlength{\belowdisplayskip}{3pt}
\begin{aligned}
\mathcal{L}_{T} = \sum_{i}\ell_{i}\ &=\lambda_{1} \sum_{i \neq j} \max \left(0, v_{i}^\top m_{j}-v_{i}^\top m_{i}+e\right) \\&+\lambda_{2} \sum_{i \neq j} \max \left(0, m_{i}^\top v_{j}-m_{i}^\top v_{i}+e\right) \\&+\lambda_{3} \sum_{i \neq j \neq k} C_{i j k}(v)\left(v_{i}^\top v_{j}-v_{i}^\top v_{k}\right) \\&+\lambda_{4} \sum_{i \neq j \neq k} C_{i j k}(m)\left(m_{i}^\top m_{j}-m_{i}^\top m_{k}\right)
\end{aligned}
\end{equation}
\noindent Here, $\lambda_{1}$ and $\lambda_{2}$ balance the impact of inter-modal ranking loss from video-to-music and music-to-video matching. And the impact of soft intra-modal structure loss within each modality is balanced by $\lambda_{3}$ and $\lambda_{4}$, respectively. $e$ is the margin for triplet loss and function  $C(\cdot)$ is defined as follows:
\begin{equation}
\setlength{\abovedisplayskip}{3pt}
\setlength{\belowdisplayskip}{3pt}
\scalebox{.98}{$
\begin{aligned}
C_{i j k}(x)=\operatorname{sign}\left(x_{i}^\top x_{k}-x_{i}^\top x_{j}\right)-\operatorname{sign}\left(\widetilde{x}_{i}^\top \widetilde{x}_{k}-\widetilde{x}_{i}^\top \widetilde{x}_{j}\right)
\end{aligned}
$}
\end{equation}

\noindent where $x_{i}$, $x_{j}$, and $x_{k}$ are the trainable features in multimodal space and $\widetilde{x}_{i}$, $\widetilde{x}_{j}$, and $\widetilde{x}_{k}$ are intra-modal features before passing through the embedding network. We first warm up the model using $\mathcal{L}_{T}$ to enable it to perform the preliminary VMR task without overfitting the noisy labels and then perform SSVMR for the rest epochs.
\vspace{-1em}
\subsection{Semi-Supervised Self-Training using R-Drop}
\label{sec:2.3}
\vspace{-0.5em}
As mentioned in~\cite{zhang2021understanding}, noisy samples usually have higher losses in the early stage. During the training process, the loss distributions of clean and noisy samples tend to obey two Gaussian distributions~\cite{EricArazo2019UnsupervisedLN}, where the loss of noise samples holds a bigger mean value. Therefore, we apply the widely used GMM to distinguish noisy samples by the loss of each input sample. We use GMM which has 2 components and fit it to the observations using the Expectation-Maximization (EM) algorithm. To be specific, we first calculate the probability $w_{i}$ that each sample belongs to the Gaussian component with bigger mean $g$, i.e., the probability of belonging to the noisy sample, as follows:
\begin{equation}
\setlength{\abovedisplayskip}{3pt}
\setlength{\belowdisplayskip}{3pt}
w_{i}=p\left(g\mid\ell_{i}\right)
\end{equation}
\noindent Then, we set a threshold $\tau$ for the probability. The original dataset $\mathcal{D}$ is divided into two parts, including a labeled set $\mathcal{X}$ and an unlabeled set $\mathcal{U}$. The label of the sample in unlabeled set $\mathcal{U}$ is most likely wrong, so we replace it by a new label. We generate the new soft label $\hat{m}$ by sharpening the predicted distribution of the model to convert unlabeled set $\mathcal{U}$ to labeled set $\hat{\mathcal{U}}$, which can encourage the model to produce lower-entropy predictions:
\setlength{\abovedisplayskip}{3pt}
\setlength{\belowdisplayskip}{3pt}
\begin{align}
\mathcal{X}&=\left\{\left(v_{i}, m_{i}\right) \mid v_{i} \in \mathcal{D}, w_{i} \leq \tau\right\} \\
\mathcal{U}&=\left\{\left(v_{i}\right) \mid v_{i} \in \mathcal{D}, w_{i}>\tau\right\}\\
\hat{m} &=\operatorname{Sharpen}(p(v, \theta), T) \\
\hat{\mathcal{U}} &=\left\{\left(v_{i}, \hat{m}_{i}\right) \mid v_{i} \in \mathcal{U}\right\} \\
\hat{\mathcal{D}} &=\mathcal{X} \cup \hat{\mathcal{U}}
\end{align}
where $p(v, \theta)$ denotes the prediction of sample $v$, Sharpen $(\cdot)$ is the temperature sharpening function commonly used in self-training and $T$ denotes the temperature:
\begin{equation}
\setlength{\abovedisplayskip}{3pt}
\setlength{\belowdisplayskip}{3pt}
\operatorname{Sharpen}(p, T)_{i}=p_{i}^{\frac{1}{T}} / \sum_{j=1}^{L} p_{j}^{\frac{1}{T}}
\end{equation}
\noindent There is an inconsistency between training and inference of the dropout based models, because dropout is used in training but not in inference, which will be more evident in noisy samples and cause more harm to the model. So we introduced R-Drop~\cite{wu2021r} loss $\mathcal{L}_{R}$ to overcome the inconsistency:
\begin{equation}
\setlength{\abovedisplayskip}{3pt}
\setlength{\belowdisplayskip}{3pt}
\begin{aligned}
\mathcal{L}_{R} =&\sum_{m \in \mathcal{U}} \frac{1}{2}\left(\mathcal{D}_{K L}\left(p_{1}(m, \theta) \| p_{2}(m, \theta)\right)\right.\\
&\left.+\mathcal{D}_{K L}\left(p_{2}(m, \theta) \| p_{1}(m, \theta)\right)\right)
\end{aligned}    
\end{equation}
where $p_{1}(m, \theta)$ and  $p_{2}(m, \theta)$ are two predicted distributions obtained by inputting the same sample twice.  $\mathcal{D}_{K L}(a \| b)$ denotes \textit{Kullback-Leibler} divergence, which plays a role in regularizing these two probability distributions. We combine the clean samples with original labels and the noisy samples with generated soft labels to obtain the new dataset $\hat{\mathcal{D}}$. 
\vspace{-1em}
\subsection{Mix at Span Level Relying on Saliency}
\label{sec:2.4}
\vspace{-0.5em}
There may be a small segment of the video that plays the most critical role in the outcome of the retrieval music. By capturing the critical video clips, it will be possible for the model to further improve performance more easily by introducing pre-training techniques~\cite{tian2023designing}. Inspired by~\cite{kim2020puzzle,yoon2021ssmix, cheng2022m3st}, to improve the ability of our framework to capture critical video information, we mix the two videos at span level. First, We compute the gradient of $\mathcal{L}_{T}$ with respect to input embedding $v$, and use its magnitude as the saliency:
\begin{equation}
\setlength{\abovedisplayskip}{3pt}
\setlength{\belowdisplayskip}{3pt}
\begin{aligned}
s=\|\partial \mathcal{L} / \partial e\|_{2}
\end{aligned}    
\end{equation}
where $\|\cdot\|_{2}$ denotes L2 norm. For two \textit{video-music} pairs denoted as $(v_1, m_1)$ and $(v_2, m_2)$, we select the least salient span $x_{1}^{S}$ from $v_1$ and replace it with the most salient span $x_{2}^{S}$ from $v_2$ to obtain a new video embedding $\hat{v}$ which is the concatenation of $\left(v_{1}^{L}; v_{2}^{S}; v_{1}^{R}\right)$ where $v_{1}^{L}$ and $v_{1}^{R}$ are embedding located to the left and the right side of $x_{1}^{S}$ respectively in the original video $x_{1}$. The selected span length of these two videos are the same and are calculated as follows:
\begin{equation}
\setlength{\abovedisplayskip}{3pt}
\setlength{\belowdisplayskip}{3pt}
\ell_{v1}=\ell_{v2}=\max \left(\min \left(\left[\lambda_{0}\left|v_{1}\right|\right],\left|v_{2}\right|\right), 1\right)    
\end{equation}
where $\ell_{v1}$ and $\ell_{v2}$ are the length and $\lambda_{0}$ is the initial mixup ratio. We calculate the triplet loss $\mathcal{L}_{T}$ of the augmented output with original video embedding of each sample and combine them by weight $\lambda$ to obtain the mix loss $\mathcal{L}_{M}$:
\begin{gather}
\setlength{\abovedisplayskip}{3pt}
\setlength{\belowdisplayskip}{3pt}
\lambda =\left|v_{2}^{S}\right| /\left|\hat{v}\right|\\
\mathcal{L}_{M}=\lambda \mathcal{L}_{T}\left( \hat{v},m_{1}) \right)+\left( 1-\lambda \right)\mathcal{L}_{T}\left( \hat{v},m_{2} \right)    
\end{gather}

\noindent With the triplet loss and mix loss, the final training objective is as follows:
\begin{equation}
\setlength{\abovedisplayskip}{3pt}
\setlength{\belowdisplayskip}{3pt}
\mathcal{L}=\mathcal{L}_{T}+ \mathcal{L}_{R} + \mathcal{L}_{M}
\end{equation}

\vspace{-1em}
\subsection{Back Retrieval}
\label{sec:2.5}
\vspace{-0.5em}
Before training video-music retrieval model as above, we conduct back retrieval to obtain more data. In the neural machine translation task, back translation~\cite{sennrich2016improving} is widely used to augment the parallel training corpus. Inspired by this, we first train a music-video retrieval model with the original dataset, and then feed each music into the model to obtain the three most similar videos. Considering that the matching criteria between video and music is more ambiguous than that of other cross-modal tasks such as image-to-text retrieval, we randomly select one of the three videos and combine it with the music into a new pair of data.

\section{Experiments}
\label{sec:exp}
\vspace{-0.5em}
\subsection{Datasets and Settings}
\vspace{-0.5em}
Following~\cite{9533662}, we train our model on HIMV-50K~\cite{9533662} dataset and evaluate it on MVD~\cite{AlexanderSchindler2016HarnessingMV} dataset. HIMV-50K is a large collection of musical videos corresponding to the part of clips of the YouTube-8M~\cite{abu2016youtube} dataset annotated as ``music video", which consists of 50,000 \textit{video-music} pairs without annotations. Note that while the audio of these clips always contains music, the video can be anything from professional promotional clips to amateur montages of still images,  i.e.\, the quality of the video part of the clips is very heterogeneous. Therefore, we use MVD~\cite{AlexanderSchindler2016HarnessingMV} dataset to evaluate our model. It consists of 2,212 music video clips of professional quality. The average duration of each clip is 4 minutes. We randomly selected 2000 of these clips to evaluate our model.

In this paper, the parameters $\lambda_{1}$, $\lambda_{2}$, $\lambda_{3}$, $\lambda_{4}$ and $e$ are set as 3, 1, 0.2, 0.2 and 6 respectively. Temperature \textit{T} is set as 0.8, threshold $\tau$ is set as 0.3 and $\lambda_{0}$ is set as 0.4. We use Adam~\cite{kingma2014adam} to optimize the parameters in our model with the learning rate $4e$-$4$ and use dropout with probability 0.9. In addition, we use Recall@K~\cite{wei2020universal} to evaluate the model performance, which is a standard metric for cross-modal retrieval that calculates the average hit ratio of the matched music clips over all the music clips that are ranked with top K match scores.

\vspace{-1em}
\subsection{Main Results}
\vspace{-0.5em}
Table \ref{tab:results} shows our experiment results on the MVD dataset and the performance of the reference VMR systems. We have the following observations: 1) Compared with baseline \textit{CMMR}~\cite{BochenLi2019QueryBV} and \textit{CBVMR}~\cite{hong2018cbvmr}, our proposed framework gain a large improvement. The reason is that \textit{SSVMR} considers the label noise which is extremely harmful to training. It's worth noticing that the parameters between \textit{SSVMR} and \textit{CBVMR} are of the same magnitude, which further verifies that the contribution of \textit{SSVMR} comes from the semi-supervised self-training using R-Drop rather than parameters factor. 2) Although \textit{CMVAE}~\cite{JingYi2021CrossmodalVA} models the matching of relevant music to video as a multimodal cross-generation problem with product-of-experts principle as the fusion strategy, \textit{SSVMR} still outperforms it. We attribute it to the fact that \textit{SSVMR} can capture critical video information and \textit{CMVAE} is proposed to retrieve music for video on the dataset, where a music clip could be associated to multiple videos. However, MVD is a dataset that associates one music clip to one video only, which can't provide sufficient information for \textit{CMVAE} to train an encoder network well. 3) \textit{MRCMV} can efficiently improve performance by leveraging the widely added voice-overs, which are the voices of unseen narrators. But when voice-overs are not added, \textit{SSVMR} outperforms it, which verifies that our framework can leverage original data better.
\renewcommand{\arraystretch}{1.05} 
\begin{table}[ht]
\vspace{-1.0em}
\centering
\begin{tabular}{lccc}
\toprule[1pt]
\textbf{Models} & \textbf{R@1} & \textbf{R@10} & \textbf{R@25} \\ \midrule[1pt]
CMMR~\cite{BochenLi2019QueryBV} & 1.65 & 6.45 & 12.70\\
CBVMR~\cite{hong2018cbvmr, 9533662} & 1.85 & 7.00 & 12.90\\
CMVAE~\cite{JingYi2021CrossmodalVA} & 2.15 & 10.95 & 16.30\\
MRCMV~\cite{li2021deep} &3.30 &12.65 &18.20\\
\midrule\midrule
\textbf{SSVMR} &\textbf{4.45} &\textbf{15.20} &\textbf{24.50}\\
\textbf{Improve} & \textbf{34.8\%} & \textbf{20.2\%} & \textbf{34.6\%}\\ 
\bottomrule[1pt]
\end{tabular}
\caption{Results on MVD datasets. Best results are in \textbf{bold}.}
\label{tab:results}
\vspace{-1em}
\end{table}
\vspace{-1em}
\subsection{Analysis}
\vspace{-0.5em}
\subsubsection{Ablation Test}
\vspace{-0.5em}
To study the effectiveness of each component in our approach, we perform ablation experiments on the MVD dataset, whose results are shown in Table \ref{tab:ablation}. We remove the back retrieval module, which is called \textit{w/o BR} in Table \ref{tab:ablation}. From the results, we can observe that the absence of back retrieval leads to 0.55\% R@1 drop, which indicates that back retrieval enlarges the dataset and encourages the robustness of our model.

\renewcommand{\arraystretch}{1.05} 
\begin{table}[!ht]
\centering
\begin{tabular}{lc}
\toprule[1pt]
\textbf{Models} & \textbf{R@1} \\ \midrule[1pt]
\textbf{SSVMR} & \textbf{4.45} \\
\midrule
\textit{w/o BR} & 3.90($\downarrow$ 0.55)\\
\textit{w/o BR + Mix} & 3.15($\downarrow$ 1.30)\\
\textit{w/o BR + Mix + R-Drop} & 2.50($\downarrow$ 1.95)\\
\textit{w/o BR + Mix + R-Drop + SL} & 1.85($\downarrow$ 2.60)\\
\bottomrule[1pt]
\end{tabular}
\caption{Ablation results on MVD.}
\label{tab:ablation}
\vspace{-1em}
\end{table}

We further remove the mix relying on saliency module, named as \textit{w/o BR + Mix} in Table \ref{tab:ablation}. We can obviously observe the R@1 drops by 1.30\%. This indicates that mixing the two videos at span level, which retains more information related to the prediction relying on saliency, can improve the ability of model to capture critical video information, allowing the model to better retrieve music.

We finally remove the R-Drop module and then remove the self-training module, denoted as \textit{w/o BR + Mix + R-Drop} and \textit{w/o BR + Mix + R-Drop + SL} respectively. The R@1 also drops, which verifies that these modules help to handle the label noise and are beneficial for the model.
\vspace{-0.5em}
\subsubsection{Effect of the number of spans replaced}
\vspace{-0.5em}
When we mix two videos at span level, only one span of the video is replaced. To check that if it's enough to only choose one span, we conduct a series of experiments shown in \ref{tab:span}. We observe that as the number of span increases, R@1 drops gradually. The reason is that replacing too many spans will break semantics and be harmful to model instead.
\renewcommand{\arraystretch}{1.05} 
\begin{table}[!ht]
\centering
\begin{tabular}{cc}
\toprule[1pt]
\textbf{N} & \textbf{R@1} \\ \midrule[1pt]
\textbf{1} & \textbf{4.45} \\
\midrule
2 & 4.27($\downarrow$ 0.18)\\
3 & 4.03($\downarrow$ 0.42)\\
4 & 3.87($\downarrow$ 0.58)\\
\bottomrule[1pt]
\end{tabular}
\caption{Results on MVD with different number of span.}
\label{tab:span}
\vspace{-1em}
\end{table}

\section{Conclusion}
\vspace{-0.5em}
In this paper, we propose \textbf{S}aliency-based \textbf{S}elf-training for \textbf{V}ideo-\textbf{M}usic \textbf{R}etrieval (\textbf{SSVMR}) to use semi-supervised self-training with R-Drop. Besides, we mix the two videos at span level relying saliency and use back retrieval to enlarge the dataset. Experiments and analysis demonstrate the effectiveness of our proposed method, which can handle the label noise, enhance the ability to capture critical video clips and enhance the robustness. In the future, we will explore how to add pre-training techniques to our approach to further improve the performance of the model.

\bibliographystyle{IEEEbib}
\bibliography{strings,refs}

\end{document}